# ASTROPHYSICS

# TWO-COMPONENT VARIABILITY OF THE SEMI-REGULAR PULSATING STAR U DELPHINI

I.L. Andronov[1], L.L.Chinarova[2]

[1] Department "High and Applied Mathematics", Odessa National Maritime University, Odessa, Ukraine, *tt_ari @ukr.net*

[2] Astronomical observatory, Odessa National University, Odessa, Ukraine, *chinarova @pochta.ru*

ABSTRACT. Photometric analysis of photometric variability of the semi-regular pulsating variable U Del is analyzed. From the international AFOEV database, 6231 brightness values in the time interval JD 2451602-55378 were chosen. For the periodogram analysis, we have used a trigonometric polynomial fit. Using the criterion of minimal variance of the approximation at arguments of observations, the optimal degree is $s=1$. Initial epoch for maximum is $T_0=JD2453340\pm3$, the period $P=1198\pm4^d$. Mean brightness at maximum is $6.624\pm0.005$, at minimum $7.124\pm0.005$, i.e. the amplitude is $0.499\pm0.005^m$. Besides this slow variability, there is a faster oscillation of a period of $119.45\pm0.06^d$, amplitude $0.303\pm0.005^m$ and an initial epoch for maximum $2453215.1\pm0.5$. These results are mean during the time interval after that analyzed in the catalogue of Chinarova and Andronov (2000). Also the method of "running sines" with a filter half-width $\Delta t=0.5P$. The local mean (averaged over a short period) brightness varies in a range $6.58-7.41^m$, the semi-amplitude exhibits very strong variations from 0.01 to $0.46^m$. The phase is also variable – typically of a full amplitude of 0.5. Close to JD 2452589, occurred a phase jump by a complete period during a descending branch of a slow wave. This effect was not observed during other cycles. No significant correlation between mean brightness and amplitude of short-period oscillation was found. Despite significant variability of amplitude, the periodic contribution is not statistically significant. Also characteristics of individual brightness extrema were found.

Long-period variables (LPVs) are pulsating stars at late stages of evolution. According to the "General Catalogue of Variable Stars" (GCVS, Kholopov et al, 1985; Samus' et al., 2012), the main types are L, M, SR, RV, which are further subdivided into subtype. A review on long-period variables was presented by Kudashkina (2003). A catalogue of characteristics of 173 variables was published by Andronov and Chinarova (2000). After that time, new intensive observations were carried out, allowing further study of these variables. Some of them (e.g. RU And) show drastic variations of an amplitude of pulsations, formally indicating switches between the types of high-amplitude Mira-type stars and low-amplitude semi-regular variables. Sometimes pulsations may practically disappear (cf. Chinarova 2010). Other similar stars (S Aql, S Tri, Y Per) were recently discussed by Marsakova and Andronov (2012).

An analysis of historical variability of U Del was presented by Thompson (1998), who argued for a main period of $1150^d$-$1210^d$. In the GCVS, $110^d$ is mentioned as the main period, type SRb, and Sp M5 II or III. From the AFOEV database, we removed "bad" and "fainter than" data, so used 6231 points. The periodogram analysis was carried out using a trigonometric polynomial fit (Andronov, 1994; Andronov and Baklanov 2004). The statistically optimal degree $s=1$. Corresponding parameters are listed in the abstract. The periodogram is shown in Fig.1.

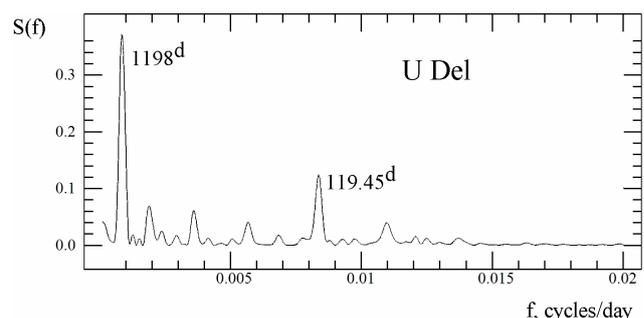

Figure 1: Periodogram $S(f)$ for U Del.

The highest peak corresponds to the "long" period $P=1198\pm4^d$. The second peak corresponds to the "short" period $119.45\pm0.06^d$. Its relatively small height is due to a low coherence.

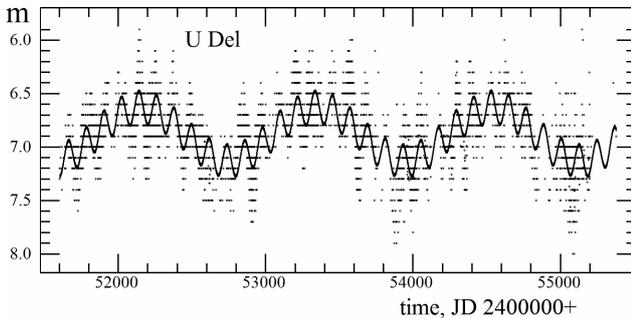

Figure 2: Individual observations and a bi-periodic fit using the periods $P=1198^d$ and $119.45^d$ computed with the program MCV (Andronov and Baklanov, 2004).

For the analysis of variability of "approximately periodic" waves, we have applied the method of "running sines" (see Andronov (1997, 2003) for a detailed theory of running approximations using arbitrary basic and weight functions). In brief, the signal is approximated as

$$m_c(t) = C_0 - R\cos(2\pi((t-T_0)/P - \phi)) \qquad (1)$$

where $C_0(t_0, \Delta t)$, $R(t_0, \Delta t)$, and $\phi(t_0, \Delta t)$ are functions of the "shift" $t_0$ and "scale" $\Delta t$ in the "wavelet terminology" (cf. Andronov, 1998). They may be interpreted as a "local mean" value (averaged over the period of "short" pulsation), semi-amplitude and phase, respectively. For our analysis, we have used a rectangular filter of a half-width $\Delta t = P/2$, so for a local approximation using the function (1), the observations from a trial interval $[t_0 - \Delta t, t_0 + \Delta t]$ are used. Also we computed envelope lines for maxima $(C_0 - R)$ and minima $(C_0 + R)$.

Contrary to a bi-periodic fit, where the amplitudes and phases are suggested to be constant, the running sine allows to check time variability of the parameters of individual cycles. Also one may see that "slow variations" ($C_0$) are not sinusoidal, one may even suggest more steep decline than incline. A noticeable phase jump at JD 2452700 by a complete period occurred at a decline, when the amplitude of pulsations drastically decreased. However, at other moments, the typical variation of phase reaches $0.5P$, thus arguing for a better effectiveness of this method as compared with a model with stable periods and amplitudes.

The semi-amplitude varies from a practical zero ($0.01^m$) to $0.42^m$. This is formally similar to RU And (Chinarova 2010), where the type of variability switched between Mira and SR (or even constant), and is listed in the GCVS as SRa. However, there were no long-period variations of the mean brightness in RU And, and in U Del the intervals of low amplitude are comparatively very short.

Thompson mentioned a drastic increase of the amplitude of variations in 1942-1949yrs, with a magnitude range of $5^m$-$8^m$. Unfortunately, his compilation of data is not available as a table of observations, thus we can't apply the same methods to provide an analysis of the all available interval of observations. This argues for a general necessity of publication of own data electronically to be available for the community.

Although the amplitude changes are significant, and some time intervals of drastic increase or decrease of amplitude were seen, the phase changes are typically smooth. There is no statistical dependence of characteristics of "fast" ($120^d$) variability with phase and/or brightness of the "slow" ($1200^d$) wave. These observational results may be used for probing models of pulsations.

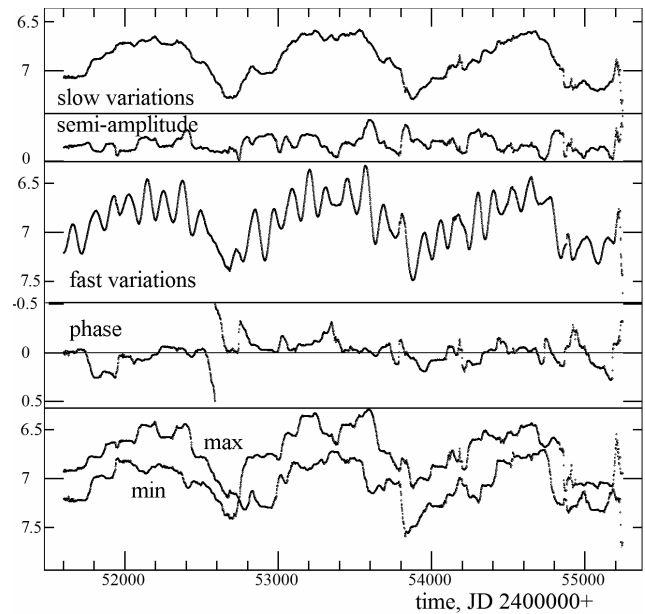

Figure 3: Characteristics of the "running sine" fit.